\documentclass[prd,aps,superscriptaddress,preprintnumbers,notitlepage,nofootinbib]{revtex4}
\usepackage{tablefootnote}
\usepackage[T1]{fontenc}
\usepackage{amsmath}
\usepackage{amssymb}
\usepackage{adjustbox}
\usepackage[mathscr]{eucal}
\usepackage{multirow}
\usepackage{hyperref}
\usepackage{xcolor}
\usepackage{braket}
\usepackage{slashed}
\usepackage{diagbox}
\usepackage[utf8]{inputenc}
\usepackage{tcolorbox}
\usepackage{tabularx}
\usepackage{bbm}
\usepackage{bm}

\newcommand{\p}[1]{$\mathbf{P}$}
\newcommand{\cp}[1]{$\mathbf{CP}$}

\begin{document}
	\baselineskip=12pt \parskip=3pt
	
	\raggedbottom
	
	\title{Large CP violation in \boldmath$\Lambda^0_b \to pK^-\pi^+\pi^-$ and its U-spin partner decays}

	\author{Xiao-Gang He}\email{hexg@sjtu.edu.cn}
	
	\affiliation{State Key Laboratory of Dark Matter Physics, Tsung-Dao Lee Institute and School of Physics and Astronomy, Shanghai Jiao Tong University, Shanghai 201210, China}
	\affiliation{Key Laboratory for Particle Astrophysics and Cosmology (MOE) \& Shanghai Key Laboratory for Particle Physics and Cosmology, Tsung-Dao Lee Institute and School of Physics and Astronomy, Shanghai Jiao Tong University, Shanghai 201210, China}

	\author{ Chia-Wei Liu }\email{chiaweiliu@sjtu.edu.cn}
	
	\affiliation{State Key Laboratory of Dark Matter Physics, Tsung-Dao Lee Institute and School of Physics and Astronomy, Shanghai Jiao Tong University, Shanghai 201210, China}
	\affiliation{Key Laboratory for Particle Astrophysics and Cosmology (MOE) \& Shanghai Key Laboratory for Particle Physics and Cosmology, Tsung-Dao Lee Institute and School of Physics and Astronomy, Shanghai Jiao Tong University, Shanghai 201210, China}
	
	\author{Jusak Tandean}\email{jtandean@yahoo.com}
	
	\affiliation{State Key Laboratory of Dark Matter Physics, Tsung-Dao Lee Institute and School of Physics and Astronomy, Shanghai Jiao Tong University, Shanghai 201210, China}
	\affiliation{Key Laboratory for Particle Astrophysics and Cosmology (MOE) \& Shanghai Key Laboratory for Particle Physics and Cosmology, Tsung-Dao Lee Institute and School of Physics and Astronomy, Shanghai Jiao Tong University, Shanghai 201210, China}
	
	\date{\today}
	
	\begin{abstract}
		The LHCb Collaboration has recently found a large CP-violating rate asymmetry in the $b$-baryon decay $\Lambda^0_b \to pK^-\pi^+\pi^-$. This is the first observation of CP violation in baryon processes, opening a new window to test its standard model (SM) origin. Many more baryon decays are expected to exhibit observable signals of CP violation.	We show that there also exists large CP violation in the U-spin partner decay mode, $\Xi^0_b \to \Sigma^+\pi^- K^+K^-$, with rate asymmetry 
		$$A_{CP}(\Xi^0_b \to \Sigma^+\pi^- K^+ K^-) = - A_{CP}(\Lambda^0_b \to p K^- \pi^+ \pi^-)~ \frac{Br(\Lambda^0_b \to p K^- \pi^+ \pi^-)}{Br(\Xi^0_b \to \Sigma^+ \pi^- K^+ K^-)}~ \frac{\tau^{\Xi_b}}{\tau^{\Lambda_b}}
		$$
		in the U-spin symmetry limit.  By neglecting a subleading contribution in the amplitudes, we obtain
		$$A_{CP}(\Lambda^0_b \to p \pi^+ \pi^- \pi^-)  =
		A_{CP}( \Xi^0_b \to \Sigma^+\pi^-  K^+ K^-)  =   -(12 \pm 3   ) \%.
		$$
		These predictions provide crucial tests for the SM.
	\end{abstract}
	
	\maketitle

	\section{Introduction}
 	
	Recently the LHCb Collaboration~\cite{LHCb:2025ray} has reported its finding of sizable CP violation in the rate asymmetry of $\Lambda^0_b \to p\pi^+ K^- \pi^-$ decay given by
	\begin{align*} & 
		A_{CP} = \frac{\Gamma - \bar \Gamma}{\Gamma + \bar \Gamma} = (2.45 \pm 0.46 \pm 0.10)\%.
	\end{align*}
 with $\Gamma$ and $\bar\Gamma$ being the rates of the particle and antiparticle modes, respectively. 
	This is the first CP violation observed in baryon decay processes. In the standard model (SM), CP violation arises from the mixing of different generations of fermions~\cite{Kobayashi:1973fv}, predicting CP violation in both meson and baryon decays. It has been a longstanding puzzle that it has been discovered only in meson decays and mixing. Therefore, the ground-breaking LHCb measurement constitutes a significant progress in the pursuit of understanding the origin of CP violation. This new result opens a new window to test the SM and search for new physics beyond it.
	
	Here we propose new baryonic decay modes with large CP violation to further test the SM. We focus on processes involving final-state mesons ($K$ and $\pi$) that are all electrically charged. The initial baryons are beauty baryons that form a doublet under U-spin symmetry: $\Lambda^0_b$ and $\Xi^0_b$. The final-state baryons also belong to a U-spin doublet: $p$ and $\Sigma^+$. We find that there exists large CP violation in the U-spin partner decay mode $\Xi^0_b \to \Sigma^+ K^+ K^- \pi^-$, with the relation
	\begin{eqnarray}
		A_{CP}(\Xi^0_b \to \Sigma^+ \pi^-K^+ K^- ) = - A_{CP}(\Lambda^0_b \to p K^- \pi^+ \pi^-) \frac{Br(\Lambda^0_b \to p K^- \pi^+ \pi^-)}{Br(\Xi^0_b \to \Sigma^+ \pi^-K^+ K^-)} \frac{\tau^{\Xi_b}}{\tau^{\Lambda_b}}\;. \label{asy}
	\end{eqnarray}
	This holds in the U-spin limit. Here, $Br $ and $\tau $ denote the branching ratio and lifetime. 	We also find large CP violation in several other related processes when subleading  contributions are neglected.
	
	Theoretical studies of four-body decays of a beauty baryon are difficult due to the nonperturbative nature of QCD at the relevant low energy scale. However, for certain processes, perturbative QCD calculations can provide reasonable guidance.
We utilize a light-quark flavor subgroup, the U-spin symmetry~\cite{Gell-Mann:1964ewy,Zweig:1964ruk} (invariance under the exchange of $d$ and $s$ quarks), as our guide~\cite{Deshpande:1994ii,He:1998rq,Gronau:2000zy,Xu:2013rua,Wang:2024rwf,He:2013vta}.
This symmetry is broken by the mass difference (several tens of MeV) between the $d$ and $s$ quarks.
However, since this mass difference is much smaller than the beauty-baryon mass scale (about 5 GeV), the breaking effects are expected to be of order $10 \sim 20\%$~\cite{He:2013vta}.
This expectation is supported by comparisons between predictions and data in beauty-meson charmless  decays into two pseudoscalar octet mesons.
For example, U-spin symmetry successfully predicts CP asymmetries in decays such as $B_d \to K^+ \pi^-$ and $B_s \to K^- \pi^+$~\cite{He:2013vta}.
	
	We should note that this approach does not predict the magnitudes of decay amplitudes.
	However, once sufficient data are collected, the relevant amplitudes can be obtained. 
	Even without all the amplitudes being determining, there exist direct relations that can guide the search for decay modes with large CP violation in four-body decays of beauty baryons, thus testing the model.
	
	The specific processes to be studied are 
	\begin{eqnarray}
		\Lambda^0_b \to p \pi^+ K^- \pi^-,\; p K^+ K^- K^-,\; \Sigma^+  K ^+ K^- \pi^-,\; \Sigma^+  K^+ \pi ^-  \pi ^-,\;  \Sigma ^+  \pi ^+ \pi^- \pi^-,\; p K^+ K^- \pi^-,\; p \pi^+ \pi^- \pi^-\;,
	\end{eqnarray}
	and their corresponding U-spin partners 
	\begin{eqnarray}
		\Xi^0_b \to \Sigma^+ K^+ K^- \pi^-,\; \Sigma^+ \pi^+ \pi^- \pi^-,\; p \pi ^+ K^- \pi^-,\; p  \pi ^+ K ^- K ^-,\; p K ^+ K^- K^-,\; \Sigma^+ \pi^+ K^- \pi^-,\; \Sigma^+  K^+ K^- K ^-\;.
	\end{eqnarray}
	U-spin symmetry implies several relations between the CP asymmetries in these partner decay channels, similarly to eq.~(\ref{asy}).
	
	For U-spin partner processes with strangeness changes $\Delta S = 0$ and $\Delta S = -1$, the amplitudes $\mathscr A^{\Delta S=0}$ and $\mathscr A^{\Delta S = -1}$ take the form:
	\begin{eqnarray}
		\mathscr A^{\Delta S = 0} = V_{ud}^* V_{ub}^{}\, \mathscr T + V_{td}^* V_{tb}^{}\, \mathscr P\;, \quad \mathscr A^{\Delta S = -1} = V_{us}^* V_{ub}^{}\, \mathscr T + V_{ts}^* V_{tb}^{}\, \mathscr P\;. \label{amplitude}
	\end{eqnarray}
	Here, $\mathscr T$ and $\mathscr P$ are the CP-conserving amplitudes for the tree and penguin contributions, respectively. 
	We then obtain the following CP-violation relation:
	\begin{eqnarray}
		A_{CP}(\Delta S = 0) = - A_{CP}(\Delta S = -1) \frac{Br(\Delta S = -1)}{Br(\Delta S = 0)} \frac{\tau^{\Delta S = 0}}{\tau^{\Delta S = -1}}\;.
	\end{eqnarray}
	When applied to $\Lambda^0_b \to pK^- \pi^+ \pi^-$ and $\Xi^0_b \to \Sigma^+ K^+ K^- \pi^-$, this gives eq.~(\ref{asy}).

	\section{Decay amplitudes}
	
	The $\Lambda_b$ baryon is a baryon that contains a heavy $b$ quark and two light quarks, $u$ and $d$, forming the $udb$ state. It is one of the light quark flavor $SU(3)_F$ triplet $b$-baryons, ($\Xi_b^0$, $-\Xi_b^-$, $\Lambda^0_b$), whose corresponding quark contents are ($dsb$, $usb$, $udb$). The U-spin symmetry group is a subgroup $SU(2)_U$ of the flavor $SU(3)_F$, taking $d$ and $s$ as the fundamental building blocks. $\Lambda^0_b$ and $\Xi_b^0$ form a doublet of U-spin: ${\bf B}_b^i = (\Lambda^0_b, \Xi^0_b)$. The upper component has $U=1/2$, $U_3=+1/2$, and the lower component has $U=1/2$, $U_3=-1/2$, while $\Xi^-_b$ is a singlet.
	
For the particles in the final states ($\Sigma^+$, $p$, $K^\pm$, and $\pi^\pm$), the proton and $\Sigma^+$ form a U-spin doublet ${\bf B}^i = (p, \Sigma^+)$, and $\pi^-$ and $K^-$ also form a U-spin doublet $(P^-)^i = (P_1, P_2) = (\pi^-, K^-)$.
For the complex conjugate of these doublets, we adopt the expressions $\overline{\bf B}_i = (\overline{p}, \overline{\Sigma^+})$ and $P_i^+  = (\pi^+, K^+)$.
	
	We concentrate on processes related to $\Lambda^0_b \to p K^- \pi^+ \pi^-$ through the U-spin symmetry. 
	The leading effective Hamiltonian responsible for these decays in the SM is given by~\cite{Buchalla:1995vs}
	\begin{eqnarray}
		\mathcal{H}_{\text{eff}} &=&  \frac{G_F}{\sqrt{2}} \left[ V_{uq}^* V_{ub} \left( C_1 Q^q_1 + C_2 Q^q_2 \right) - V_{tq}^*V_{tb} \sum_{i=3}^{10} C_i Q^q_i \right] + \text{H.c.}, \notag \\
		Q^q_1 &=& (\bar u \gamma^\mu (1 - \gamma_5) u) (\bar q \gamma_\mu (1 - \gamma_5) b), \quad
		Q^q_2 = (\bar q \gamma^\mu (1 - \gamma_5) u) (\bar u \gamma_\mu (1 - \gamma_5) b),
	\end{eqnarray}
	where $q = d,s$, $G_F$ is the Fermi constant, $C_{1,2}$ are the Wilson coefficients for the tree-level operators, $C_{3,...,10}$ are the Wilson coefficients for the strong and electroweak penguin operators, and $V_{ij}$ are the elements of the Cabibbo-Kobayashi-Maskawa (CKM) matrix.

	For $q = d$, $\mathcal{H}_{\text{eff}}$ has U-spin quantum numbers $U=1/2$, $U_3=+1/2$. We use $H_i = (1, 0)$ to indicate this, with the understanding that the CKM elements are to be added to the amplitudes it induces. Similarly, for $q = s$, we use $H_i = (0,1)$. The decay amplitudes are formed by taking the matrix elements $M = \langle {\bf B} P^+P^-_1(p_1) P^-_2(p_2)|\mathcal{H}_{\text{eff}}| {\bf B}_b \rangle$ with $p_i$ the momenta. Removing the KM elements from the decay amplitudes, we can write
	\begin{eqnarray}
		\mathscr T &=&   a_T  \overline{\bf B}_i {\bf B}_b^i {\cal H}_T^j (P^-)^k (P^+_1 P^+_2)_{\{jk\}}
		+   b_T \overline{\bf B}_i {\bf B}_b^k {\cal H}_T^j (P^-)^i (P^+_1 P^+_2)_{\{jk\}}   +
		c_T \overline{\bf B}_i {\bf B}_b^k {\cal H}_T^i (P^-)^j (P^+_1 P^+_2)_{\{jk\}}
		\;,
	\end{eqnarray}
	where
	$
	(P_1^+ P_2^+)_{\{ij\}} = \left[ (P_1^+)_i (P_2^+)_j + (P_1^+)_j (P_2^+)_i \right] / 2,
	$
	and a symmetry factor of 2 needs to be added if $P_1^+ = P_2^+$. In general, $a_T$, $b_T$, and $c_T$ have momentum dependence on $p_i$, and the above form ensures that the amplitudes are symmetric under the exchange of two identical particles in the final state.
	
	Similarly, one can define the penguin amplitudes ($\mathscr P$) by replacing  the subscripts $T$ with $P$ in the foregoing. 
	Expanding the above amplitudes, we have the U-spin partner pair decay amplitudes given by
	\begin{eqnarray}
		\label{partner-pairs}
		1) &&\mathscr A(\Lambda^0_b \to p K^-\pi^+ \pi^-) = V_{us}^*V_{ub}(a_T + b_T) + V^*_{ts}V_{tb}(a_P + b_P)\;,\nonumber\\
		&&\mathscr A(\Xi^0_b \to \Sigma^+ \pi^-K^+ K^- ) = V_{ud}^*V_{ub}(a_T + b_T) + V^*_{td}V_{tb}(a_P + b_P)\;;\nonumber\\
		&&\nonumber\\
		2) &&\mathscr A(\Lambda^0_b \to p K^+ K^- K^-) = 2V_{us}^*V_{ub}a_T  + 2 V^*_{ts}V_{tb}a_P\;,\nonumber\\
		&&\mathscr A(\Xi^0_b \to \Sigma^+ \pi^+ \pi^-\pi^-) = 2 V_{ud}^*V_{ub}a_T +2  V^*_{td}V_{tb} a_P\;;\nonumber\\
		&&\nonumber\\
		3) &&\mathscr A(\Xi^0_b \to p \pi ^+ \pi ^- K ^-) = V_{us}^*V_{ub}(b_T+c_T) + V^*_{ts}V_{tb}(b_P + c_P) \;,\nonumber\\
		&&\mathscr A(\Lambda^0_b \to \Sigma^+ K ^+ K^- \pi^- ) = V_{ud}^*V_{ub}(b_T+c_T) + V^*_{td}V_{tb}(b_P+c_P)\;;\nonumber\\
		&&\nonumber\\
		4) &&\mathscr A(\Xi^0_b \to p \pi^+ K^- K ^-) = 2 V_{us}^*V_{ub}b_T + 2  V^*_{ts}V_{tb}b_P \;,\nonumber\\
		&&\mathscr A(\Lambda^0_b \to \Sigma^+ K^+\pi ^- \pi ^-) = 2 V_{ud}^*V_{ub}b_T +2  V^*_{td}V_{tb}b_P\;; \\
		&&\nonumber\\
		5) &&\mathscr A(\Lambda^0_b \to \Sigma^+ \pi^+ \pi ^- \pi ^-) = 2 V_{us}^*V_{ub}c_T +2  V^*_{ts}V_{tb} c_P\;,\nonumber\\
		&&\mathscr A(\Xi^0_b \to p K^+ K^- K ^-) = 2 V_{ud}^*V_{ub} c_T+ 2 V^*_{td}V_{tb} c _P\;;\nonumber\\
		&&\nonumber\\
		6) &&\mathscr A(\Xi^0_b \to \Sigma^+ \pi^+ K^- \pi^-) = V_{us}^*V_{ub}(a_T+c_T) + V^*_{ts}V_{tb}(a_P + c_P) \;,\nonumber\\
		&&\mathscr A(\Lambda^0_b \to p K^+K^-\pi^-) = V_{ud}^*V_{ub}(a_T+c_T) + V^*_{td}V_{tb}(a_P+c_P)\;;\nonumber\\
		&&\nonumber\\
		7) &&\mathscr A(\Xi^0_b \to \Sigma^+ K^+ K^- K^-) =2  V_{us}^*V_{ub}(a_T+b_T+c_T) + 2 V^*_{ts}V_{tb}(a_P + b_P+c_P) \;,\nonumber\\
		&&\mathscr A(\Lambda^0_b \to p \pi^+ \pi^-\pi^-) = 2 V_{ud}^*V_{ub}(a_T+b_T+c_T) + 2 V^*_{td}V_{tb}(a_P+b_P+c_P)\;. \nonumber   
	\end{eqnarray}
	In the above, we have omitted the $p_1$ and $p_2$ labeling with the understanding that when there are two identical particles in the final state, the amplitude should be symmetrized. Each of the pairs has the form in eq.~(\ref{amplitude}), and therefore similar CP asymmetry relations exist as in eq.~(\ref{asy}). 
	
	\section{ Discussions and conclusions}
	
	As already mentioned earlier, U-spin symmetry can predict relations between U-spin partner decay channels, as shown in eq.~\eqref{partner-pairs} and the following
	\begin{eqnarray}\label{CPV-relation}
		1) &&	A_{CP}(\Xi^0_b \to \Sigma^+ K^+ K^- \pi^-) =- A_{CP}(\Lambda^0_b \to p \pi^+ K^- \pi^-) {Br(\Lambda^0_b \to p  \pi^+ K^- \pi^-) \over  Br(\Xi^0_b \to \Sigma^+ K^+ K^-\pi^-)}{\tau^{\Xi_b}\over \tau^{\Lambda_b}}\;,\nonumber\\
		2) &&	A_{CP}(\Xi^0_b \to \Sigma^+ \pi^+\pi^- \pi^-) =- A_{CP}(\Lambda^0_b \to p K^+ K^- K^-) {Br(\Lambda^0_b \to p  K^+ K^- K^-) \over  Br(\Xi^0_b \to \Sigma^+ \pi^+ \pi^-\pi^-)}{\tau^{\Xi_b}\over \tau^{\Lambda_b}}\;,\nonumber\\
		3) &&	A_{CP}(\Lambda^0_b \to \Sigma^+ K ^+ K^- \pi^-) =- A_{CP}(\Xi^0_b \to p \pi ^+ K^- \pi^-) {Br(\Xi^0_b \to p  \pi ^+ K^- \pi^-) \over  Br(\Lambda^0_b \to\Sigma^+ K ^+ \pi^- \pi^-)}{\tau^{\Lambda_b}\over \tau^{\Xi_b}}\;,\nonumber\\
		4) &&	A_{CP}(\Lambda^0_b \to \Sigma^+ K^+\pi^- \pi^-) =- A_{CP}(\Xi^0_b \to p \pi^+ K^- K^- ) {Br(\Xi^0_b \to p \pi^+ K  ^-K^-) \over  Br(\Lambda^0_b \to\Sigma^+ K^+\pi^- \pi^-)}{\tau^{\Lambda_b}\over \tau^{\Xi_b}}\;, \\
		5) &&	A_{CP}(\Xi^0_b \to p K^+ K^-K ^-) =- A_{CP}(\Lambda_b \to\Sigma^+  \pi^+ \pi ^-\pi ^-) {Br(\Lambda^0_b \to \Sigma^+  \pi^+ \pi ^-\pi ^- ) \over  Br(\Xi^0_b \to p K^+K^- K^- )}{\tau^{\Xi_b}\over \tau^{\Lambda_b}}\;,\nonumber\\
		6) &&	A_{CP}(\Lambda^0_b \to p K^+  K^- \pi^-) =- A_{CP}(\Xi^0_b \to \Sigma^+ \pi^+ K^- \pi^-) {Br(\Xi^0_b \to \Sigma^+  \pi^+ K^- \pi^-) \over  Br(\Lambda^0_b \to p K^+ K^- \pi^-)} {\tau^{\Lambda_b}\over \tau^{\Xi_b}}\;,\nonumber\\
		7) &&	A_{CP}(\Lambda^0_b \to p \pi^+ \pi^- \pi^-) = - A_{CP}(\Xi^0_b \to \Sigma^+ K^+ K^- K^-) {Br(\Xi^0_b \to \Sigma^+  K^+ K^- K^-) \over  Br(\Lambda^0_b \to  p \pi^+ \pi^-\pi^-)} {\tau^{\Lambda_b}\over \tau^{\Xi_b}}\;. \nonumber
	\end{eqnarray}
	From the above, we clearly see that if the branching ratios and lifetimes are known, then using one of the measured CP violations in a decay channel, the CP violation in the other channels can be predicted.
	
	We have the following related experimental information about the lifetimes and branching ratios~\cite{ParticleDataGroup:2024cfk}
	\begin{eqnarray}
		&&\tau^{\Lambda^0_b} = (1.47\pm 0.09)\times 10^{-12} \text{s}\;,\;\;\tau^{\Xi^0_b} = (1.480\pm 0.03)\times 10^{-12} \text{s}\;; \nonumber\\
		&&Br(\Lambda^0_b \to p \pi^+\pi^-\pi^-) = (2.08\pm 0.21)\times 10^{-5}\;,\;\;Br(\Lambda^0_b \to p K^+ K^- \pi^-) = (4.0\pm0.6)\times 10^{-6}\;,  \\
		&&Br(\Lambda^0_b \to p \pi^+K^-\pi^-) = (5.0\pm 0.5)\times 10^{-5}\;,\;\;Br(\Lambda^0_b \to p K^+ K^- K^-) = (1.25\pm 0.13)\times 10^{-5}\;. \nonumber
	\end{eqnarray}
For $\Xi^0_b$ decays, the experimental measurements are
$
	 {Br}(\Xi^0_b \to p K^+ K^- K^-)\, \mathcal{R}  = (1.7 \pm 0.9) \times 10^{-7}, 
	 {Br}(\Xi^0_b \to p \pi^+ K^- K^-)\, \mathcal{R} = (1.7 \pm 0.3) \times 10^{-6}, $ and  $ 
	 {Br}(\Xi^0_b \to p \pi^+ K^- \pi^-)\, \mathcal{R}  = (1.9 \pm 0.4) \times 10^{-6} 
$~\cite{ParticleDataGroup:2024cfk}.
Here, $\mathcal{R} =  {Br}(b \to \Xi_b)/ {Br}(b \to \Lambda_b)$, and $ {Br}(b \to \mathbf{B}_b)$ denotes the fraction of a $b$ quark hadronizing into a baryon $\mathbf{B}_b$.
The hierarchy  of 
$
 {Br}(\Xi^0_b \to p \pi^+ K^- \pi^-) \approx  {Br}(\Xi^0_b \to p \pi ^+ K^- K^-) \gg  {Br}(\Xi^0_b \to p K ^+ K^- K^-)
$   
implies that   \(   a_i ,  b_i  \gg c_i \).

	Efforts for CP violation searches have also been made. Besides the recently measured CP violation
	$A_{CP} (\Lambda_b \to p \pi^+ K^- \pi^-)= (2.45\pm0.46\pm 0.10)\%$,
	for other decay modes discussed here, the precision is at the level of a few percent, but no confirmed results have been reported.
	
	From 1) in Eq.~(\ref{CPV-relation}), we would obtain:
	\begin{eqnarray}
		A_{CP}(\Xi^0_b \to \Sigma^+ K^+ K^- \pi^-) =-  (2.45 \pm 0.47) \times 10^{-2} {(5.0 \pm 0.5 )\times 10^{-5}\over Br(\Xi^0_b \to \Sigma^+ K^+ K^- \pi^-)}\;.
	\end{eqnarray}
	Unfortunately, experimentally the branching ratio $Br(\Xi^0_b \to \Sigma^+ K^+ K^- \pi^-)$ has not been measured.  
	If
	$Br(\Xi^0_b \to \Sigma^+ K^+ K^- \pi^-)$ is similar to $Br(\Lambda^0_b \to p K^- \pi^+ \pi^-)$, then sizeable CP violation may arise, potentially measurable in the near future.
	
As mentioned earlier, the sizes of  $c_i$ are much smaller than  those of  $a_i$ and $b_i$.  By taking the subleading contribution of $c_i$ to be zero,  more predictions can be made. In this case,
we would have~\cite{ParticleDataGroup:2024cfk}
	\begin{eqnarray}
		&& 
		2
\frac{  \tau^{\Lambda^0_b}} {\tau^{\Xi^0_b} }
Br(\Xi^0_b \to \Sigma^+ K^+ K^- \pi^-)   =	Br(\Lambda^0_b \to p \pi^+ \pi^- \pi^-) = 
 (2.08\pm 0.21
		)\times 10^{-5}\;, \nonumber\\
		&&
		\frac{1}{2}
\frac{  \tau^{\Lambda^0_b}} {\tau^{\Xi^0_b} }
Br(\Xi^0_b \to \Sigma^+ K^+ K^- K^-)  = Br(\Lambda^0_b \to p \pi^+ K^- \pi^-)  = (5.0\pm 0.5  )\times 10^{-5}\;,\\
		&&
A_{CP}(\Lambda^0_b \to p \pi^+ \pi^- \pi^-)
=
 - A_{CP}(\Lambda^0_b \to p \pi^+ K^- \pi^-) {2 Br(\Lambda^0_b \to p  \pi^+ K^- \pi^-) \over  Br(\Lambda^0_b \to p \pi^+ \pi^- \pi^- )} =   ( -12\pm 3    ) \%\;,\nonumber 
	\end{eqnarray}
	and
	$A_{CP}(\Xi^0_b \to \Sigma^+ K^+ K^- \pi^-) = A_{CP}(\Lambda^0_b \to p \pi^+ \pi^- \pi^-)$ from 1) and 7). 
The experimental measurement  
$A_{CP}(\Lambda_b^0 \to p \pi^+ \pi^- \pi^-) = (1.1 \pm 2.6)\%$~\cite{ParticleDataGroup:2024cfk}  
is consistent with zero at the current stage. Nevertheless, the
deviation  in 
 magnitude suggests a potential tension and warrants further experimental investigation.
	We strongly encourage a renewed experimental investigation of
$A_{CP}(\Lambda_b^0 \to p \pi^+ \pi^- \pi^-)$.
	On the other hand, 2) and 6) lead to
	\begin{eqnarray}
		&&
Br(\Lambda^0_b \to p K^+ K^- K^-) =		2 \frac{  \tau^{\Lambda^0_b}} {\tau^{\Xi^0_b} }
		Br(\Xi^0_b \to \Sigma^+ \pi^+ K^- \pi^-) =   (1.25\pm 0.13
		)\times 10^{-5}\;,\nonumber\\
		&&
Br(\Lambda^0_b \to p K^+ K^- \pi^-) =		\frac{1}{2}\frac{  \tau^{\Lambda^0_b}} {\tau^{\Xi^0_b} }
		Br(\Xi^0_b \to \Sigma^+ \pi^+ \pi^- \pi^-) =   (4.0\pm0.6
		)\times 10^{-6}\;, \\
		&&
		- A_{CP}(\Lambda^0_b \to p K^+ K^- K^- ) {  	
			Br(\Lambda^0_b \to p K^+ K^- K^- )\over 2
			Br(\Lambda^0_b \to p K^+ K ^- \pi^-)}  
		= 
	A_{CP}(\Lambda^0_b \to p K^+ K ^- \pi^-)
		= (-0.3 \pm  3.0      ) \% \,,\nonumber 
	\end{eqnarray}
	which is consistent with the experimental value of $( -7 \pm 5 )\%$.
	The above
	 predictions   can be tested by experiments in the near future.
	
	In the above, we have worked in the limit of U-spin symmetry. As already mentioned, U-spin symmetry in $B_{d,s}$ decays into two pseudoscalar mesons holds to about 20\%. We would also expect a similar level of robustness for the above predictions. To obtain CP asymmetries experimentally, one should avoid using data points in resonant regions, since small U-spin violations may be magnified near such regions.

	\begin{acknowledgments}
		This work is supported in part by
		the National Key Research and Development Program of China under Grant No. 2020YFC2201501, by
		the Fundamental Research Funds for the Central Universities, by National Natural Science Foundation of P.R. China (No.12090064, 12205063, 12375088 and W2441004). 
		J.T. thanks the Tsung-Dao Lee Institute, Shanghai Jiao Tong University, for hospitality and kind support during the
		preparation of this paper.\\
	\end{acknowledgments}
	
	{\it Note added}:
	While we were finishing our work, a preprint by Zhang and Wang appeared~[arXiv:2503.21885], which also considers the
	U-spin for related decays.

\end{document}